\documentclass[aps,pre,twocolumn,superscriptaddress,longbibliography,floatfix]{revtex4-2}
\usepackage{graphicx,bm,amssymb,amsmath,dcolumn,hyperref}
\usepackage{multirow}
\usepackage{enumerate}
\usepackage{enumitem}
\usepackage{color}
\usepackage{epstopdf}
\usepackage{bbm}
\usepackage{hyperref}
\usepackage{threeparttable}
\usepackage{amsthm}
\usepackage{bm}        
\usepackage{mathtools}
\usepackage{color}
\usepackage{tikz}
\usepackage{float}
\usepackage{empheq}
\usepackage[ruled,linesnumbered]{algorithm2e}
\definecolor{blue}{rgb}{0,0,1}
\definecolor{red}{rgb}{1,0,0}
\definecolor{green}{rgb}{0,1,0}

\newcommand{\PPa}{
\begin{tikzpicture}[baseline=1pt]
 \draw [fill] (0,0) circle(0.20ex);
 \draw [fill] (0.25,0) circle(0.20ex);
 \draw [fill] (0,0.25) circle(0.20ex);
 \draw [fill] (0.25,0.25) circle(0.20ex);
 \draw [thick] (0,0)--(0,0.25);
 \draw [thick] (0.25,0)--(0.25,0.25);
\end{tikzpicture}
}

\newcommand{\PPb}{
\begin{tikzpicture}[baseline=1pt]
 \draw [fill] (0,0) circle(0.20ex);
 \draw [fill] (0.25,0) circle(0.20ex);
 \draw [fill] (0,0.25) circle(0.20ex);
 \draw [fill] (0.25,0.25) circle(0.20ex);
 \draw [thick] (0,0)--(0.25,0);
 \draw [thick] (0,0.25)--(0.25,0.25);
\end{tikzpicture}
}

\begin{document}

\title{Sweeny dynamics for the random-cluster model with small $Q$}

\author{Zirui Peng }
\affiliation{Department of Modern Physics, University of Science and Technology of China, Hefei, Anhui 230026, China}

\author{Eren Metin Elçi}
\affiliation{School of Mathematical Sciences, Monash University, Clayton, VIC 3800, Australia}
\author{Youjin Deng}
\email{yjdeng@ustc.edu.cn}
\affiliation{Department of Modern Physics, University of Science and Technology of China, Hefei, Anhui 230026, China}
\affiliation{Hefei National Laboratory, University of Science and Technology of China, Hefei 230088, China}
\affiliation{MinJiang Collaborative Center for Theoretical Physics, College of Physics and Electronic Information Engineering, Minjiang University, Fuzhou 350108, China}
\author{Hao Hu}
\email{huhao@ahu.edu.cn}
\affiliation{School of Physics and Optoelectronic Engineering, Anhui University, Hefei, Anhui 230601, China}

\begin{abstract}
The Sweeny algorithm for the $Q$-state random-cluster model in two dimensions is shown to exhibit a rich mixture of critical dynamical scaling behaviors. As $Q$ decreases, the so-called critical speeding-up for non-local quantities becomes more and more pronounced. However, for some quantity of specific local pattern---e.g., the number of half faces on the square lattice, we observe that, as $Q \to 0$, the integrated autocorrelation time $\tau$ diverges as $Q^{-\zeta}$, with $\zeta \simeq 1/2$, leading to the non-ergodicity of the Sweeny method for $Q \to 0$. Such $Q$-dependent critical slowing-down, attributed to the peculiar form of the critical bond weight $v=\sqrt{Q}$, can be eliminated by a combination of the Sweeny and the Kawasaki algorithm. Moreover, by classifying the occupied bonds into bridge bonds and backbone bonds, and the empty bonds into internal-perimeter bonds and external-perimeter bonds, one can formulate an improved version of the Sweeny-Kawasaki method such that the autocorrelation time for any quantity is of order $O(1)$. 
\end{abstract}
\maketitle

\section{\label{sec1}Introduction }
The Hamiltonian of the $Q$-state Potts model~\cite{wupotts} reads $\mathcal{H}=-J\sum _{\langle ij\rangle}\delta_{\sigma_i,\sigma_j}$, where $J>0$ is the ferromagnetic coupling constant, $\langle ij\rangle$ stands for a pair of nearest-neighboring sites, and each lattice site $i$ has a spin $\sigma_i$, taking one of the $Q$ states $\sigma_i=1,2,...,Q$. For $Q=2$, the Potts model corresponds to the famous Ising model. Given a spin configuration, under the Fortuin-Kasteleyn transformation~\cite{kasteleyn1969phase,fortuin1972random},  bonds between nearest-neighboring sites with the same spin value are independently occupied with the probability $p=1-e^{-J/kT}$ ($k$ is the Boltzmann constant, $T$ is the temperature, and, for convenience, we set $kT=1$), 
and the resulting graph is a configuration of the random-cluster model~\cite{grimmett2006random}.
For bond configurations $A\subseteq{E}$ of a given graph $G=(V,E)$, the random-cluster model is defined by the partition function
\begin{align}
    Z=\sum_{A\subseteq{E}} Q^{k(A)}v^{|A|} \;.
\end{align}
Here $A$ is the set of ``occupied bonds", $|A|$ is the number of occupied bonds, $k(A)$ is the number of clusters (connected components), real positive numbers $Q$ and $v=p/(1-p)$ are the cluster and bond weights, respectively.
For integer $Q\ge2$, the random-cluster model serves as a graphical representation of the $Q$-state Potts model.
As $Q \to 1$, the random-cluster model reduces to independent bond percolation~\cite{stauffer1992introduction}. 

{In equilibrium, the weight ratio between two configurations are $W(\beta) / W(\alpha) = Q^{\delta k(A)} v^{\delta |A|} $, where $\delta k(A)$ ($\delta |A|$) represents the changes of $k(A)$ ($|A|$) from configuration $\alpha$ to $\beta$. As $Q \rightarrow 0$, when $v/Q$ is finite, a configuration consists of a spanning forest~\cite{jacobsen2005spanning,grimmett2006random}. This is because in this case $W(\beta) / W(\alpha) \sim Q^{\delta [k(A)+|A|]} $, and spanning forests have the minimum $k(A)+|A|$. From a spanning forest $\alpha$, occupying any internal-perimeter bond (defined later in Fig. 2) leads to a nonforest $\beta$ with $\delta[k(A)+|A|]=1$, such that $W(\beta) / W(\alpha) \sim Q \rightarrow 0$. In the case $Q \rightarrow 0$ with $v/Q^{\sigma}(0< \sigma <1)$ being finite, similarly the model consists of uniform spanning trees which have the smallest $k(A)+\sigma |A|$~\cite{jacobsen2005spanning,grimmett2006random}.}

Markov Chain Monte Carlo (MCMC) algorithms~\cite{newman1999monte,landau2005guide} are effective in simulating the random-cluster model, such as the Swendsen-Wang-Chayes-Machta algorithm~\cite{swendsen1987nonuniversal,chayes1998graphical} for $Q>1$ with non-local cluster updates, and the Sweeny algorithm~\cite{sweeny1983monte} for $Q>0$ with local bond updates. To assess the efficiency of the MCMC algorithms, autocorrelation times are commonly used as a metric~\cite{sokal1997functional}. For any observable ${\cal O}$, one first defines the autocorrelation function 
\begin{align}
    \rho_{\cal O}(t) = \frac{\langle {{\cal O}_{s}{\cal O}_{s+t}}\rangle-{\langle {\cal O}\rangle}^2}{{\rm var}({\cal O})} \;.
\end{align}
Here expectations are taken after equilibrium is attained, and the time unit is one ``hit", in which only one bond is visited (a Monte Carlo sweep consists of $|E|$ hits, where $|E|$ is the number of all bonds). Then {one defines} the exponential autocorrelation time for an observable $\cal O$
\begin{align}
    \tau_{{\rm exp},\cal O} = \limsup_{t \rightarrow \pm \infty}\frac{|t|}{{\rm-log}|\rho_{\cal O}(t)|} \;,
\end{align}
and the overall exponential autocorrelation time
\begin{align}
    \tau_{{\rm exp}}=\sup_{\cal O}\tau_{{\rm exp},\cal O} \;.
\end{align}
The integrated autocorrelation time for an observable $\cal O$ is defined as
\begin{align}
    \tau_{{\rm int},\cal O}=\frac{1}{2}\sum_{t=-\infty}^{\infty}\rho_{\cal O}(t) \;.
\end{align}

Observables typically have $\tau_{{\rm exp},{\cal O}}=\tau_{\rm exp}$ and $\tau_{{\rm int},\cal O} \le \tau_{\rm exp}$. 
The autocorrelation times $\tau_{\rm exp}$ and $\tau_{{\rm int},\cal O}$ play different roles in MCMC simulations~\cite{sokal1997functional}: $\tau_{{\rm exp}}$ is the relaxation time of the slowest mode in the system, and $\tau_{{\rm int},\cal O}$ controls the statistical error in the measurement of $\langle \cal O \rangle$. Thus {$\tau_{{\rm int},\cal O}$} mainly determines the efficiency of MCMC sampling.
 For a time series of total length $M$, the sample mean
\begin{align}
    \widehat{{\cal O}}=\frac{1}{M}\sum_{t=1}^{M}{\cal O}_t
\end{align}
has variance
\begin{align}
    {\rm var}(\widehat{{\cal O}}) \approx 2\tau_{{\rm int},{\cal O}} \frac{{\rm var}({\cal O})}{M},    {\; \rm when \;} M\to \infty \;.
    \label{Eq:measureTau}
\end{align}
The above Eq.~(\ref{Eq:measureTau}) can be used to measure $\tau_{{\rm int},{\cal O}}$.

In MCMC simulations, the dynamic processes {usually} undergo {\it critical slowing-down}~\cite{hohenberg1977theory}: 
the autocorrelation time diverges as $\tau/{|E|} \sim \xi^z$ when the critical point is {being} approached, 
where $\xi$ is the spatial correlation length and $z$ is {called the} dynamic exponent. For two-dimensional Ising model, 
Metropolis-Hastings~\cite{metropolis1953equation,hastings1970monte} and Glauber~\cite{glauber1963time,landau2005guide} algorithms have $z\approx 2$. 
~To overcome the $critical$ $slowing$-$down$, the Swendsen-Wang-Chayes-Machta algorithm~\cite{swendsen1987nonuniversal,chayes1998graphical} was proposed for the random-cluster model with $Q>1$. 
The latter algorithm has $z\approx 0.2$ for the two-dimensional Ising model, 
which represents a huge improvement. 
The gained efficiency is associated with nonlocal cluster updates.

\begin{figure}[t]
    \centering
    \includegraphics[width=0.5\textwidth]{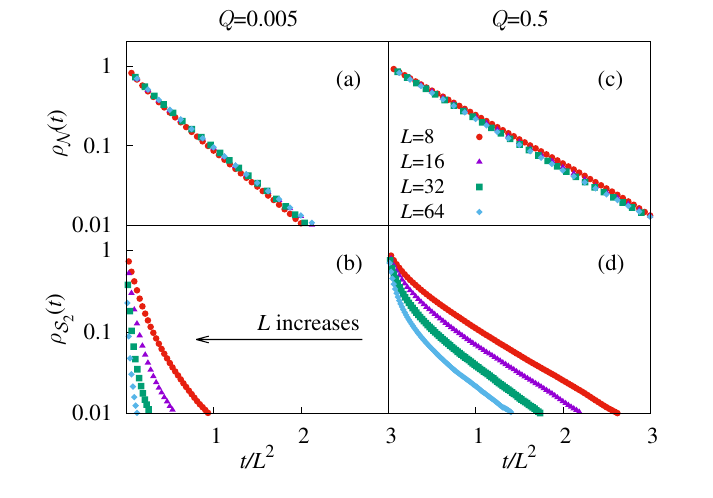}
    \caption{Autocorrelation function $\rho_{\mathcal{S}_2}(t)$ and $\rho_{\mathcal{N}}(t)$ for the critical 2D random-cluster model at $Q=0.005$ and $Q=0.5$ on the square lattice. The unit of $t$ is one hit.}
    \label{S2N}
\end{figure}

However, some local algorithms may still be very efficient, such as the worm algorithm~\cite{prokof2001worm} and the Sweeny algorithm~\cite{deng2007critical}.
The worm algorithm outperforms the cluster algorithms for the three-dimensional Ising model in measuring quantities such as the susceptibility and correlation length.
With local single-bond updates, the Sweeny algorithm can surprisingly exhibit {\it critical speeding-up} for {a suitable range of} $Q$~\cite{deng2007critical}:
the relaxation time of some observables decreases significantly at time scales much shorter than one sweep as the lattice size $L$ increases, and this phenomenon becomes even more pronounced as $Q$ becomes smaller. We illustrate the latter property of the Sweeny algorithm by observing the dynamical behavior of two quantities, i.e., the number of occupied bonds $ \mathcal{N} = |A|$ and the second moment of cluster sizes $\mathcal{S}_2 = \sum{|\mathcal{C }|^2}$.  
Hereafter, if not specified, all simulations were performed on the square lattice at the critical point $v_c(Q)=\sqrt{Q}$.
The average $\langle \mathcal{S}_2 \rangle / |V|$ is a susceptibility-like quantity of the random-cluster model~\cite{kasteleyn1969phase,fortuin1972random}, with $|V|$ being the volume. 
For $Q \ge 2$, it is found~\cite{deng2007critical} that the dynamic exponent $z_{exp} \approx z_{{\rm int},\mathcal{N}}$, and that both $z_{{\rm int},\mathcal{N}}$ and $z_{{\rm int},\mathcal{S}_2}$ are only slightly larger than the lower bound $\alpha / \nu$~\cite{li1989rigorous}, and could possibly be equal to it (here $\alpha$ and $\nu$ are critical exponents for the specific heat and correlation length, respectively). 
For $Q < 2$, the autocorrelation function of $\mathcal{N}$ follows an almost purely exponential function, as demonstrated for $Q=0.005$ and $Q=0.5$ in Fig.~\ref{S2N}(a) and ~\ref{S2N}(c), respectively. 
However, the dynamic behavior of $\mathcal{S}_2$ is richer~\cite{deng2007critical}: (1) as shown in Fig.~\ref{S2N}(b) and \ref{S2N}(d), the autocorrelation function $ \rho_{\mathcal{S}_2}(t)$ decays very rapidly in much less than one sweep, as a prelude to the final exponential decay $\sim e^{-t / \tau_{exp}}$;
(2) the exponent $z_{{\rm int},\mathcal{S}_2}$ is a negative number, and it decreases with the increasing of lattice size $L$;
(3) from Fig.~\ref{S2N}(d) to \ref{S2N}(b), it is seen that the critical speeding-up is more pronounced for smaller $Q$.

\begin{figure}[t]
\centering
\includegraphics[width=0.5\textwidth]{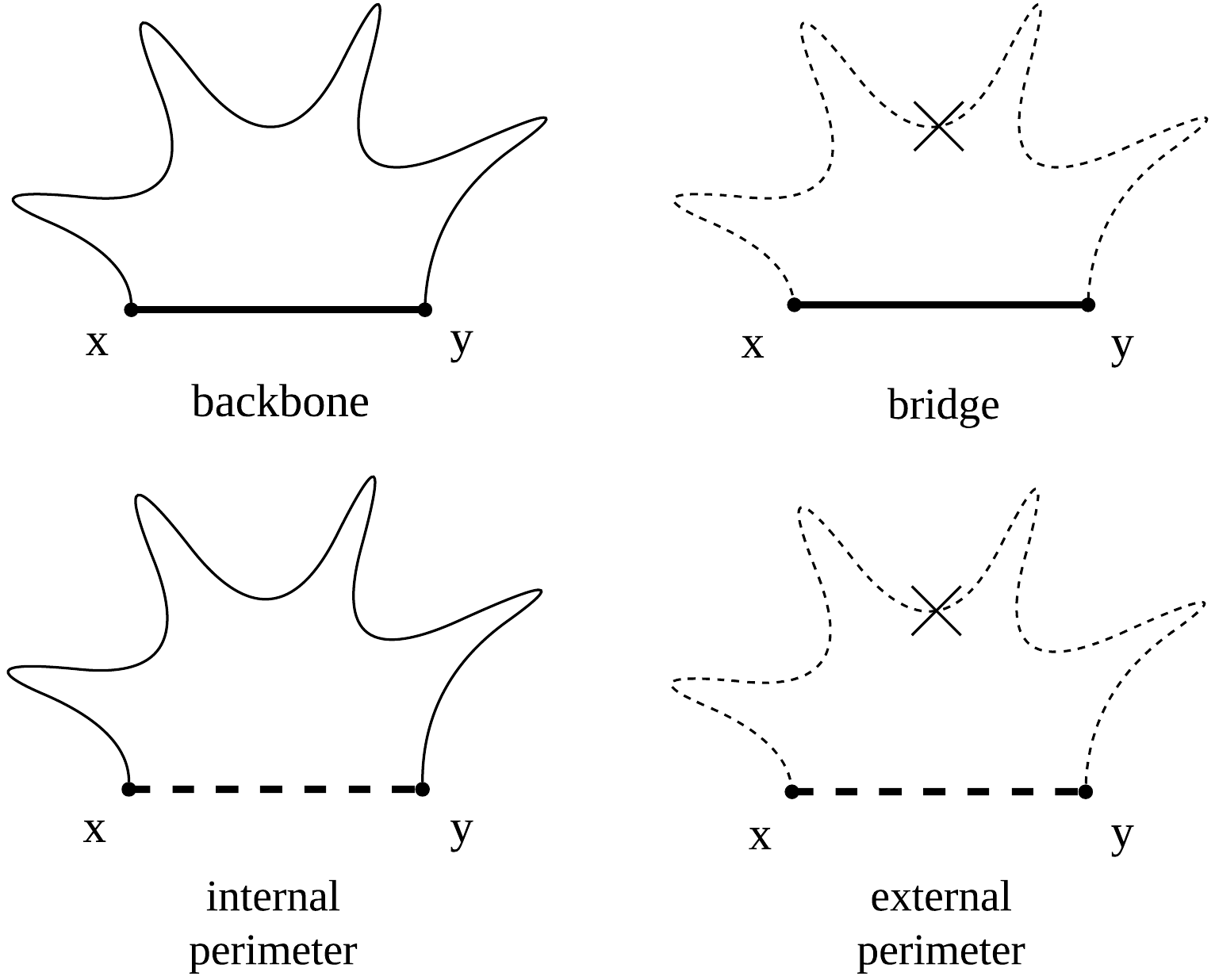}
\caption{Schematic diagrams for four kinds of bonds 
in the random-cluster model. They are categorized according to their occupation and connectivity: if bond $xy$ is occupied and the two endpoints are connected (resp. not connected) via a path that does not include $xy$, then the occupied bond is called a $backbone$ $bond$ (resp. $bridge$ $bond$); if bond $xy$ is empty and the two endpoints are connected (resp. not connected) via a path that does not include $xy$, then the empty bond is referred to as an $internal$-$perimeter$ $bond$ (resp. $external$-$perimeter$ $bond$).}
\label{bonds}
\end{figure}

In this article, we are interested in the dynamic behavior of the Sweeny algorithm for very small $Q$ values.
A $Q$-dependent critical slowing-down is found
for the random-cluster model in two dimensions. The phenomenon can be observed in the dynamics of
a specific local pattern on the square lattice, i.e., a unit square with configuration $\PPa$ or $\PPb$, where solids lines represent occupied bonds. We call the local pattern as a {\it half face.} 
We find that, when $Q\rightarrow 0$, the integrated autocorrelation time of the number of half faces in terms of sweeps (i.e., {$\tau_{{\rm int},\mathcal{N}_{\rm HF}}/2L^2$}, and hereafter $\tau_{\mathcal{N}_{\rm HF}}/L^2$ is used for brevity) diverges  as $Q^{-1/2}$.
This indicates that the Sweeny algorithm becomes not ergodic as $Q \to 0$. 

The above $Q$-dependent critical slowing-down can be explained by the acceptance probability of updates in the Sweeny algorithm. 
{As illustrated in Fig.~\ref{bonds}, considering different occupation and connectivity, the bonds can be classified into 
four types --- the $backbone$, $bridge$, $internal$-$perimeter$ and $external$-$perimeter$ bonds. 
At $v_c=\sqrt{Q}$, when transforming an internal-perimeter bond into a backbone bond, one has $\delta k(A) = 0$ and $\delta |A| = 1$, 
which yields the weight ratio $W(\beta) / W(\alpha)=\sqrt{Q}$; and transforming a bridge bond into an external-perimeter bond
leads to $\delta k(A) = 1$ and $\delta |A| = -1$, which also gives $W(\beta) / W(\alpha) = \sqrt{Q}$. It is shown in Sec.~\ref{sec2}
that these weight ratios are associated with vanishing acceptance probabilities $~\sqrt{Q}$ as $Q \rightarrow 0$.}
To overcome {the $Q$-dependent slowing-down},
we first use the knowledge of four kinds of bonds in the detail balance condition and propose an improved Sweeny algorithm. 
The improved Sweeny algorithm largely alleviates the slowing-down but does not fully solve the problem.
For example, at $Q=0$ where the graphs become spanning trees, the improved Sweeny algorithm is still invalid
due to zero probability of deleting or adding a bond. 
We then employ the Kawasaki algorithm~\cite{kawasaki1966diffusion}, which proposes the random exchange of an empty and an occupied bond. 
At $Q=0$ on spanning trees, the Kawasaki algorithm still has an $L$-dependent divergence
$\tau_{\mathcal{N}_{\rm HF}}/L^2 \sim L^{3/2}$ which can also be related to the acceptance probability of updates.
The latter divergence can be eliminated through an improved Kawasaki algorithm which uses the categories of bonds. 
Finally, we combine the use of the improved Sweeny and improved Kawasaki algorithms, and find no critical slowing-down for small $Q$, i.e., 
the autocorrelation time $\tau/L^2$ is of order ${\cal O}(1)$, being independent of $Q$.

{The remainder of this paper is organized as follows. Section~\ref{sec2} describes the Sweeny algorithm and the $Q$-dependent slowing-down. Section~\ref{sec3} presents an improved Sweeny algorithm which can greatly improve simulation efficiency. Section~\ref{sec4} introduces an improvement of the Kawasaki algorithm and its combination with the improved Sweeny algorithm, which can completely eliminate  the $Q$-dependent slowing-down. Section~\ref{sec:level1} contains a discussion and conclusion.}

\section{\label{sec2}Sweeny algorithm and $Q$-dependent slowing-down}

\begin{algorithm}[b]
\caption{Original Sweeny Algorithm}
\label{alg:sweeny0}
Choose a bond $xy$ uniformly at random\\
\uIf {$xy$ is an external-perimeter bond} 
{Occupy $xy$ with probability $\min \{1,v/Q\}$}
\uElseIf {$xy$ is an internal-perimeter bond}
{Occupy $xy$ with probability $\min\{1,v\}$}
\uElseIf {$xy$ is a bridge bond} 
{Delete $xy$ with probability $\min\{1,Q/v\}$}
\ElseIf {$xy$ is a backbone bond}
{Delete $xy$ with probability $\min\{1,1/v\}$}
\end{algorithm}

As mentioned in the Introduction, the Sweeny algorithm is a local single-bond update algorithm for the random-cluster model. Actually, it is the only known MCMC algorithm for the model with $0 <Q < 1$. 
The original Sweeny algorithm~\cite{sweeny1983monte} is presented in Alg.~\ref{alg:sweeny0} . A basic step of the Sweeny algorithm is proposing to change the status of a randomly selected bond, and accepting the proposal according to the Metropolis criterion.  

For an efficient implementation of the Sweeny algorithm, the main challenge is in checking the connectivity between the two endpoints of a chosen bond. There are three typical methods with different asymptotic run-time scaling: breadth-first search, union-and-find, and dynamic connectivity algorithms~\cite{elcci2013efficient,elci2015algorithmic}. We use the method of simultaneous breadth-first searches starting at both endpoints of a bond $xy$, since it has very simple code and is effective enough for small and medium size $L$~\cite{deng2010some}. We note that a simple modification of Alg.~\ref{alg:sweeny0} can save some time by not requiring connectivity checking for a fraction of updates. The modification for $Q\le 1$ at $v_c=\sqrt{Q}$ is shown in Alg.~\ref{alg:sweeny1}.
The basic idea is to consider together several probabilities in~Alg.~\ref{alg:sweeny0}: if the generated random number is less than the minimum of these probabilities, the update must be accepted, and if it is greater than the maximum of these probabilities, the update must be rejected. For $Q \to 1$, this small change in the algorithm can lead to a significant improvement of efficiency.

\begin{algorithm}[b]
\caption{Modified Sweeny Algorithm (for $Q \leq 1$ at $v_c=\sqrt{Q}$)}
\label{alg:sweeny1}
$P_{\rm min}$=$\sqrt{Q}$, $P_{\rm max}$=1\\
Choose a bond $xy$ uniformly at random\\
Generate a random number $P \in [0,1)$\\
\uIf {$P<=P_{\rm min}$}
{Change the status of $xy$}
\ElseIf{$xy$ is an external-perimeter bond or a backbone bond}
{Change the status of $xy$}
\end{algorithm}

As described in the Introduction, the Sweeny algorithm exhibits rich dynamics, including both critical slowing-down and critical speeding-up. However, its dynamics at small $Q$ values are not uncovered fully. For Alg.~\ref{alg:sweeny0} and Alg.~\ref{alg:sweeny1}, one can observe that 
they encounter a problem when $Q$ is small: at  $v_c=\sqrt{Q}$,
probabilities of some updates (occupying an internal-perimeter bond and deleting a bridge bond) are proportional
to $\sqrt{Q}$, which causes the algorithm being not ergodic at $Q \to 0$. 
{The problem was not severe in previous studies, since the studied $Q$ values are not extremely small, 
and conventional observables such as $\mathcal{N}$ and $\mathcal{S}_2$ could not detect any slowing-down effect due to 
the $\sqrt{Q}$-form updating probabilities.}
Actually on the square lattice in the limit $Q \to 0$, a critical equilibrium configuration consists of a spanning tree, 
thus the above observables remain fixed as $\mathcal{N}=L^2-1$ and $\mathcal{S}_2=L^4$ for a given $L$.

\begin{figure}[t] 
\centering
\includegraphics[width=0.5\textwidth]{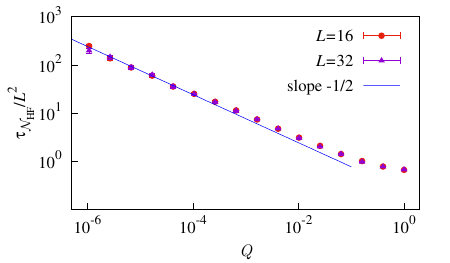}
\caption{Autocorrelation time of the number of half faces for the original Sweeny algorithm. $\tau_{\mathcal{N}_{\rm HF}}/L^2$ diverges as $Q^{-1/2}$ when $Q \to 0$. }
\label{tauNHF:sweeny}
\end{figure}

We find that the above problem can become dominant in sampling other quantities, such as the number of half faces $\mathcal{N}_{\rm HF}$.
Using the original Sweeny algorithm, we calculated $\tau_{\mathcal{N}_{\rm HF}}$ as plotted in Fig.~\ref{tauNHF:sweeny}. It can be seen that $\tau_{\mathcal{N}_{\rm HF}}/L^2$ diverges as $Q^{-1/2}$, independent of the system size $L$. We refer to this as {\it $Q$-dependent critical slowing-down}.
This newly discovered phenomenon at small $Q$ enriches the dynamics of the Sweeny algorithm.
It is important to note that, away from $Q=0$, the original Sweeny algorithm is not inherently flawed for small $Q$. 
For example, when $Q=10^{-6}$, the autocorrelation time $\tau_{\mathcal{N}_{\rm HF}}/L^2$ is of  order $O(10^2)$, which means that $\mathcal{N}_{\rm HF}$ can still be measured rather efficiently.

To understand the $Q$-dependent critical slowing-down, we associate the above factor $Q^{-1/2}$ to the updating probabilities in the Sweeny algorithm. In the following we measure the numbers of the aforementioned four kinds of bonds, and
then give these updating probabilities.

\begin{figure}[t] 
\centering
\includegraphics[width=0.5\textwidth]{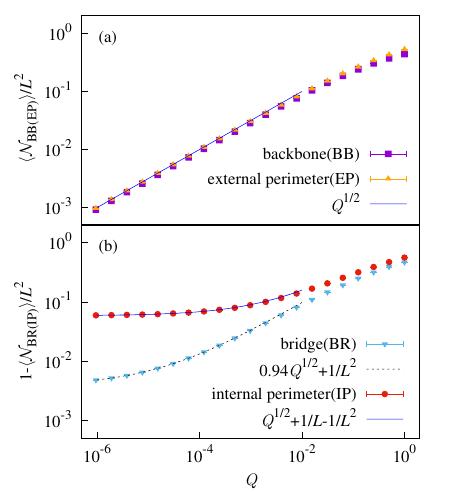}
\caption{Average numbers of four types of bonds versus $Q$ at fixed $L=16$. 
The $Q^{1/2}$-behavior in (a) is confirmed by fits in Table~\ref{table1}, and the curves in (b) can be derived from Eq.~(\ref{dfbond1}) in the main text.}
\label{bonddf}
\end{figure}

On the square lattice in the limit $Q \to 0$, the number of bridge bonds ($\mathcal{N}_{\rm BR}$) and of internal-perimeter bonds ($\mathcal{N}_{\rm IP}$) are $\mathcal{N}_{\rm BR} \to L^2-1$ and $\mathcal{N}_{\rm IP} \to |E|- \mathcal{N}_{\rm BR}$, respectively. 
These numbers can be obtained from the characteristics of a spanning tree: since every node (i.e., site) corresponds to an occupied bond except the root, there are $L^2-1$ occupied bonds for a tree with $L^2$ nodes; the $L^2-1$ occupied bonds are all bridge bonds, and other (empty) bonds are internal-perimeter bonds. For finite small $Q$, changes of the numbers of four kinds of bonds with the decrease of $Q$ are shown in Fig.~\ref{bonddf}, with fitting results for two of them in Table \ref{table1}. 
For $L=16$, we find that for small $Q$, the average number of backbone bonds ($\mathcal{N}_{\rm BB}$) and external-perimeter bonds ($\mathcal{N}_{\rm EP}$) are respectively
\begin{align}
    &\langle\mathcal{N}_{\rm BB}\rangle \simeq 0.94L^2\sqrt{Q} \;,\quad   \langle\mathcal{N}_{\rm EP}\rangle \simeq L^2\sqrt{Q} \;,
\label{dfbond0}
\end{align}
both being proportional to $\sqrt{Q}$, as plotted in Fig.~\ref{bonddf}(a). 
The average number of occupied bonds is found to be $\langle \mathcal{N}_B \rangle \simeq L^2 - 1$ for small $Q$, which is consistent with $\mathcal{N}_{\rm BR} \to L^2-1$ in the limit $Q \to 0$. In the simulations, for convenience of a loop construction which will be described in Section~\ref{sec3}, we use periodic boundary conditions in one direction of the square lattice and open boundary conditions in the other direction. The total number of bonds on the lattice is $|E|=2L^2-L$. Thus, considering $\mathcal{N}_B=\mathcal{N}_{\rm BB}+\mathcal{N}_{\rm BR}$ and $|E|-\mathcal{N}_B=\mathcal{N}_{\rm EP}+\mathcal{N}_{\rm IP}$, we derive
\begin{align}
    &\langle\mathcal{N}_{\rm BR}\rangle \simeq L^2 - 0.94L^2\sqrt{Q} -1 \simeq L^2(1 - 0.94\sqrt{Q}) \;, \nonumber \\
    &\langle\mathcal{N}_{\rm IP}\rangle \simeq L^2 - L^2\sqrt{Q} - L + 1 \simeq L^2 (1-\sqrt{Q}) \;.
\label{dfbond1}
\end{align}
The above scaling behaviors are demonstrated by a plot of $1-\langle \mathcal{N}_{\rm BR(IP)}\rangle /L^2$ as shown in Fig.~\ref{bonddf}(b).
In the thermodynamic limit $L \rightarrow \infty$, from the self-duality of the square lattice, the critical numbers of
four kinds of bonds can be obtained~\cite{Eren2016} (an alternative derivation is given in the Appendix of the present paper) as
\begin{align}
    \langle\mathcal{N}_{\rm BB}\rangle/L^2 = \langle\mathcal{N}_{\rm EP}\rangle/L^2 = \frac{\sqrt{Q}}{1+\sqrt{Q}} \;, \nonumber \\
    \langle\mathcal{N}_{\rm BR}\rangle/L^2 = \langle\mathcal{N}_{\rm IP}\rangle/L^2 = \frac{1}{1+\sqrt{Q}} \;.
\label{rhobonds}
\end{align}
In the limit $Q \rightarrow 0$, the factor $0.94$ in Eqs.~(\ref{dfbond0}) and (\ref{dfbond1}) deviates from $1$ in Eq.~(\ref{rhobonds}) due to finite-size effects.

With the above bond numbers for small $Q$, we can get the updating probabilities in the Sweeny algorithm as following:
the probability of selecting a bridge bond or internal-perimeter bond is $\sim 1$, and the acceptance probability ($Q/v$ or $v$) is $\sqrt{Q}$, then the associated updating probability is $\sim \sqrt{Q}$; 
the probability of selecting a backbone bond or external-perimeter bond is $\sim \sqrt{Q}$, and the acceptance probability is 1, then the associated updating probability is also $\sim \sqrt{Q}$. 
Thus the $Q^{-1/2}$-form critical slowing-down is associated with the $\sqrt{Q}$-form updating probabilities.

\begin{table*}
\caption{\label{table1}
Fitting results for $\langle \mathcal{N}_{\rm EP} \rangle /L^2$ and $\langle \mathcal{N}_{\rm BB} \rangle /L^2$ at $L=16$. The data were fitted by the formula $Q^{y_Q}(a_0+b_1Q^{y_Q}+b_2Q^{2y_Q})$ 
using the least-square method, with an upper cutoff $Q_{\rm max}$ on included values of $Q$.
}
\begin{ruledtabular}
\begin{tabular}{lllllll}
 
&\quad$Q_{\rm max}$  	&\quad$y_Q$ 	&\quad$a_0$ 	&\quad$b_1$ 	&\quad$b_2$	&\ 
$\chi^2/{\rm DF}$ 	\\
\hline 
&0.015625       &0.50001(8)	&0.9959(8) 	&-0.940(10)	&0.762081(1)	&10.2/11\\ 

$\langle \mathcal{N}_{\rm EP} \rangle /L^2$  &0.0078125     &0.5000(1) 	&0.996(1)  	&-0.95(2)  	&0.803890(1)	&10.1/10\\ 

&0.00390625     &0.5000(1) 	&0.996(2)  	&-0.95(3)  	&0.840794(1)	&10.0/9\\ 

&0.00195313     &0.5002(2) 	&0.999(2)  	&-1.02(6)  	&1.775890(1)	&8.0/8\\ 
\cline{1-7}
&0.015625     &0.49989(8)	&0.9404(8) 	&-0.969(10)	&0.796009(1)	&14.7/11\\ 

$\langle \mathcal{N}_{\rm BB} \rangle /L^2$ &0.0078125     &0.4999(1) 	&0.941(1)  	&-0.97(2)  	&0.815144(1)	&14.6/10\\ 

&0.00390625     &0.5001(1) 	&0.942(2)  	&-1.02(3)  	&1.284402(1)	&11.3/9\\ 

&0.00195313    &0.4999(2) 	&0.940(2)  	&-0.95(6)  	&0.401849(1)	&9.5/8\\ 
\end{tabular} 
\end{ruledtabular}
\end{table*}

\section{\label{sec3}An improved Sweeny algorithm}

To overcome the $Q$-dependent critical slowing-down, we introduce an improved Sweeny algorithm as shown in Alg.~\ref{alg:isweeny}.
Instead of selecting a bond at random, the improved algorithm randomly proposes one of the following four operations: deleting a backbone bond ({\sffamily delete-backbone}), deleting a bridge bond ({\sffamily delete-bridge}),
adding a backbone bond ({\sffamily add-backbone}), and adding a bridge bond ({\sffamily add-bridge}). 
Acceptance probabilities of the operations shall be given later.
The algorithm consists of recording and dynamically updating the information of four kinds of bonds, which are directly used in proposing MCMC updates.
To the end of this section, we shall see that the improved Sweeny algorithm can largely solve the $Q$-dependent critical slowing-down.

\begin{algorithm}[b]
\caption{Improved Sweeny Algorithm}
\label{alg:isweeny}
 Randomly choose an operation:
 {\sffamily delete-backbone}, {\sffamily delete-bridge}, {\sffamily add-backbone}, or {\sffamily  add-bridge}\\
\uIf {{\sffamily delete-backbone}}
{Randomly pick up a backbone bond and delete it with probability $P^{(-)}_{BB}$}
\uElseIf {{\sffamily delete-bridge}}
{Randomly pick up a bridge bond and delete it with probability $P^{(-)}_{BR}$}
\uElseIf {{\sffamily add-backbone}}
{Randomly pick up an internal-perimeter bond and occupy it (i.e., add a backbone bond) with probability $P^{(+)}_{BB}$}
\ElseIf{{\sffamily  add-bridge}}
 {Randomly pick up an external-perimeter bond and occupy it (i.e., add a bridge bond) with probability $P^{(+)}_{BR}$}
\end{algorithm}
\medskip

\begin{figure}[t]
\centering
\includegraphics[width=0.3\textwidth]{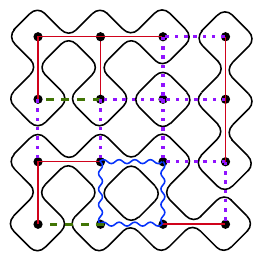}
\caption{Example bond configuration and corresponding Baxter-Kelland-Wu loops on the square lattice. Red lines are bridge bonds, blue wavy lines are backbone bonds, green dashed lines represent internal-perimeter bonds and purple dotted lines stand for external-perimeter bonds.} 
\label{loop}
\end{figure}

For Alg.~\ref{alg:isweeny}, a list is created to dynamically keep the information for each kind of bonds.
The lists record the indices of corresponding bonds, and the numbers of four types of bonds, i.e., $\mathcal{N}_{\rm BB},\mathcal{N}_{\rm BR},\mathcal{N}_{\rm EP},\mathcal{N}_{\rm IP}$.
The bonds are classified by using the Baxter-Kelland-Wu loops~\cite{baxter1976equivalence} on the medial lattice, as exemplified in Fig.~\ref{loop}. 
The loops are constructed by assigning each occupied bond two loop arcs to its two sides, and assigning each unoccupied bond two loop arcs which cross the bond.
The types of bonds are determined as following:
an occupied bond is a backbone (bridge) bond iff the loop arcs on its two sides do not (do) belong to the same loop, and an unoccupied bond is an external (internal) -perimeter bond if the two arcs crossing the bond do not (do) belong to the same loop. To avoid the case that an occupied bond is a backbone bond but has its two loop arcs in the same loop, the boundary conditions are chosen as periodic in one direction and open in the other direction such that the finite-$L$ square lattice is planar.
During the update of a bond state, the connectivity checking and labelling of loop arcs in the loop configuration are performed 
by simultaneous breadth-first searches starting from two loop arcs of the bond~\cite{deng2010some,elcci2013efficient}. 
When a loop is split into two or two loops are merged into one during the update, the simultaneous searching time is only proportional to the size of the smaller loop.

The acceptance probabilities in Alg.~\ref{alg:isweeny} are derived as follows.
Using {$\alpha$ and $\beta$} to represent two configurations that can transform into each other in an update, the detailed balance condition has the form
\begin{align}
&\pi(\alpha){P_{pro}}(\alpha\rightarrow \beta){P_{acc}}(\alpha \rightarrow \beta)\\ \nonumber
=&\pi(\beta){P_{pro}}(\beta\rightarrow \alpha){P_{acc}}(\beta \rightarrow \alpha) \;,
\end{align}
where $\pi$ is the configuration weight, $P_{pro}$ and $P_{acc}$ are the proposal and acceptance probabilities, respectively.
The detailed balance condition associated with the deletion or addition of a backbone bond can be written as
\begin{align}
    1 \cdot \frac{1}{\mathcal{N}_{\rm BB}(\alpha)} \cdot P^{(-)}_{BB} = \frac{1}{v} \cdot \frac{1}{\mathcal{N}_{\rm IP}(\beta)} \cdot P^{(+)}_{BB} \;, 
\end{align}
and that associated with the deletion or addition of a bridge bond as
\begin{align}
    1 \cdot \frac{1}{\mathcal{N}_{\rm BR}(\alpha)} \cdot P^{(-)}_{BR} = \frac{Q}{v} \cdot \frac{1}{\mathcal{N}_{\rm EP}(\beta)} \cdot P^{(+)}_{BR} \;.
\end{align}
Following the Metropolis-Hastings criterion, the acceptance probabilities are calculated as
\begin{gather}
     P^{(-)}_{BB} =\min \{ 1,\frac{\mathcal{N}_{\rm BB}(\alpha)}{\mathcal{N}_{\rm IP}(\beta)} \cdot \frac{1}{v} \} \;,\label{eq:accDeletingBB} \\
     P^{(+)}_{BB} =\min \{ 1,\frac{\mathcal{N}_{\rm IP}(\beta)}{\mathcal{N}_{\rm BB}(\alpha)} \cdot v \} \;,  \\
     P^{(-)}_{BR} =\min \{ 1,\frac{\mathcal{N}_{\rm BR}(\alpha)}{\mathcal{N}_{\rm EP}(\beta)} \cdot \frac{Q}{v} \} \;, \label{eq:accDeletingBR} \\
     P^{(+)}_{BR} =\min \{ 1,\frac{\mathcal{N}_{\rm EP}(\beta)}{\mathcal{N}_{\rm BR}(\alpha)} \cdot \frac{v}{Q} \} \;. \label{eq:accAddingBR}
\end{gather}

\begin{figure}[t] 
\centering
\includegraphics[width=0.5\textwidth]{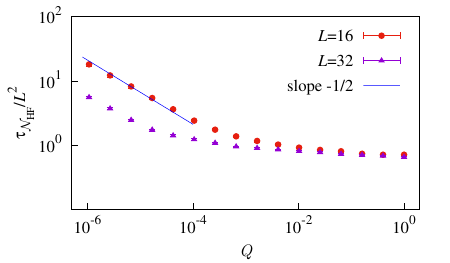}
\caption{$\tau_{\mathcal{N}_{\rm HF}}$ for the improved Sweeny algorithm.}
\label{tauNHF:isweeny}
\end{figure}

Our results of $\tau_{\mathcal{N}_{\rm HF}}$ at $v_c=\sqrt{Q}$ for the improved Sweeny algorithm is shown in Fig.~\ref{tauNHF:isweeny}. 
When fixing the size $L$, it can be seen that, as $Q$ decreases, $\tau_{\mathcal{N}_{\rm HF}}/L^2$ is almost independent of $Q$ before a turning point $Q_c(L)$,
after which $\tau_{\mathcal{N}_{\rm HF}}/L^2$ tends to increase as $\sim Q^{-1/2}$.
The value of $Q_c(L)$ decreases with increasing system size $L$, and it is expected that $Q_c(L) \rightarrow 0$ in the limit $L\rightarrow \infty$. 
Thus, comparing with results of the original Sweeny algorithm (i.e., Fig.~\ref{tauNHF:sweeny}), the improved Sweeny algorithm greatly improves simulation efficiency. 
For $Q=10^{-6}$ and $L=32$, the autocorrelation time for the improved algorithm is of order $O(10)$, and it decreases with $L$, making the algorithm sufficient for most studies of the random-cluster model with small $Q$. 

The performance of the improved Sweeny algorithm can be understood from the acceptance probabilities: 
$P^{(+)}_{BB}$ and $P^{(-)}_{BR}$ are  of order $O(1)$ for $Q \gg Q_c(L)$; $P^{(-)}_{BB}$ and $P^{(+)}_{BR}$ are of order $O(1)$ for all $Q$. 
The residual $Q^{-1/2}$ behavior is attributed to that $P^{(+)}_{BB}$ and $P^{(-)}_{BR}$ are  of order $O(\sqrt{Q})$ for $Q \ll Q_c(L)$.
Detailed analyses of these acceptance probabilities will be presented in Section~\ref{sec4}.
In the limit $Q \rightarrow 0$, the improved Sweeny algorithm fails since $P^{(-)}_{BR}=P^{(+)}_{BB}=0$, i.e., no bridge bond can be deleted from a uniform spanning tree and no backbone bond can be added to the tree.

\section{\label{sec4}An improved Kawasaki algorithm and its combination with the improved Sweeny algorithm}

In the limit $Q \rightarrow 0$, since both the original and improved Sweeny algorithms are invalid, one has to employ other methods.
When $Q \rightarrow 0$, though no bond can be erased or occupied, bonds may exchange their states. Thus, to simulate models with $Q \rightarrow 0$, we can use the Kawasaki algorithm~\cite{kawasaki1966diffusion}, in which a randomly selected empty bond is proposed to switch with a randomly selected occupied bond.

\begin{figure}[t] 
\centering
\includegraphics[width=0.5\textwidth]{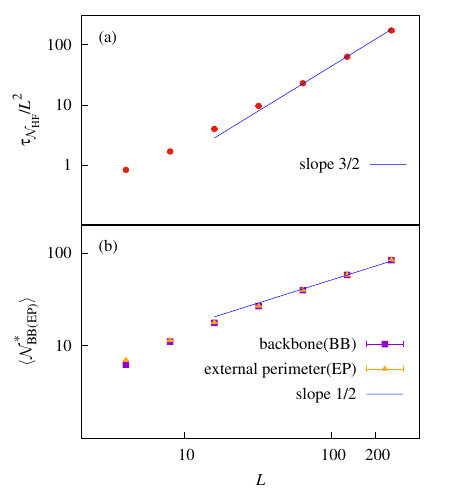}
\caption{Results for the uniform spanning trees on the square lattice.
(a) The autocorrelation time of the Kawasaki algorithm $\tau_{\mathcal{N}_{\rm HF}}/L^2$ versus $L$. Here each swap of two bond states is regarded as the time unit ``hit".
(b) The average numbers of bonds in the intermediate state $\langle\mathcal{N}^*_{\rm EP} \rangle$ and $\langle\mathcal{N}^*_{\rm BB}\rangle$ versus $L$. 
Deviations at small sizes represent finite-size corrections.
}
\label{kawasaki}
\end{figure}

We applied the Kawasaki algorithm to the uniform spanning trees,
which appear in the limit of $Q \rightarrow 0$ with $v_c/Q^{1/2}$ being finite, for the critical square-lattice model.
The results are shown in Fig.~\ref{kawasaki}(a), where it can be seen that $\tau_{\mathcal{N}_{\rm HF}}/L^2$ diverges as $\sim L^{3/2}$. 
We understand this $L$-dependent slowing-down as follows.

A Kawasaki update can be performed in two steps: first deleting a randomly chosen occupied bond so that the spanning tree is split into two trees, then occupying a randomly chosen empty bond.
We call the configuration between the two steps an intermediate state. For a uniform spanning tree, the overall update would be accepted if and only if the chosen empty bond is an external-perimeter bond in the intermediate state.

The number of external-perimeter bonds in the intermediate states $\mathcal{N}^*_{\rm EP}$ can be counted by observing changes of the loops: on a spanning tree, the deletion of an occupied bond decompose a large loop into two loops, and $\mathcal{N}^*_{\rm EP}$ is proportional to the size of the external perimeter associated with the smaller loop.
The loop surrounding a cluster defines its hull, and the fraction of loop arcs in the hull corresponding to external-perimeter bonds is called the external perimeter of the cluster.
The size of the smaller loop is determined as $\sim L^{d_{\rm H} - x_2}$ in Ref.~\cite{deng2010some}, where $d_{\rm H}$ is the fractal dimension of the hull and $x_2$ is the two-arm exponent~\cite{saleur1987exact}.
Analogously, we conjecture that the size of the external perimeter associated with the smaller loop is $\sim L^{d_{\rm EP} - x_2}$, where $d_{\rm EP}$ is the fractal dimension of the external perimeter.
According to the literature~\cite{saleur1987exact,nienhuis1984critical,cardy1998number,deng2010some}, the fractal dimensions take values $d_{\rm H}=1+2/g$ and $d_{\rm EP}=1+g/8$; the parameter $g$ and $k$-arm exponents $x_k$ are given by 
\begin{align}
   &Q=2+2\cos(g \pi /2) \;, \; g \in [2,4] \;,\\
   &x_k=(g/8)k^2-(g-4)^2/(8g) \;, \; k \geq 2 \;.
\end{align}
For $Q \rightarrow 0$,  one gets $g=2$, $x_2=3/4$, and $d_{\rm EP}=5/4$. Then we expect  $\langle \mathcal{N}^*_{\rm EP} \rangle \sim L^{5/4-3/4} \sim L^{1/2}$, which is numerically confirmed in Fig.~\ref{kawasaki}(b). 
Thus, the probability that a randomly selected empty bond is an external-perimeter bond is $\langle \mathcal{N}^*_{\rm EP} \rangle /2L^2 \sim L^{-3/2}$,
which leads to $\tau_{\mathcal{N}_{\rm HF}}/L^2 \sim L^{3/2}$ on the uniform spanning trees.

To eliminate the above $L$-dependent slowing-down, similar to the idea in improving the Sweeny algorithm, we make use of
the real-time information of four kinds of bonds and propose an improved Kawasaki algorithm, as shown in Alg.~\ref{alg:ikawasaki}. 
This improved algorithm either exchange a bridge bond with an external-perimeter bond,
or exchange a backbone bond with an internal-perimeter bond.
Since the proposed update always exist and the acceptance probability is $1$, the improved Kawasaki algorithm does not
have the above $L$-dependent slowing-down. 
This is confirmed by our simulation results which have {$\tau_{\mathcal{N}_{\rm HF}}/L^2 \simeq 0.55$ for $L = 4$ to $128$.}

\begin{algorithm}[t]
\caption{Improved Kawasaki Algorithm}
\label{alg:ikawasaki}
 Randomly choose an occupied bond and erase the bond, then we have an intermediate configuration\\
\uIf {The chosen bond is a bridge bond}
{Randomly occupy an external-perimeter bond in the intermediate configuration}
\ElseIf {The chosen bond is a backbone bond}
{Randomly occupy an internal-perimeter bond in the intermediate configuration}
\end{algorithm}
\medskip

For nonzero small $Q$, since updates in the (improved) Kawasaki algorithm cannot change the total number of occupied bonds,
one has to employ the (improved) Sweeny algorithm. In Section~\ref{sec3}, for the improved Sweeny algorithm at fixed $L$, we have shown 
the appearance of a $Q$-dependent slowing-down after an turning point $Q_c(L)$. This can be understood by observing 
the acceptance probabilities as follows.
First, for updates of deleting a backbone bond or adding a bridge bond, substituting Eqs.~(\ref{dfbond0}) and (\ref{dfbond1}) into
Eqs.~(\ref{eq:accDeletingBB}) and (\ref{eq:accAddingBR}), one has
\begin{align}
     P^{(-)}_{BB} &=\min \{ 1,\frac{\mathcal{N}_{\rm BB}(\alpha)}{\mathcal{N}_{\rm IP}(\alpha)+1} \cdot \frac{1}{\sqrt{Q}} \} \simeq 1 \;, \\
     P^{(+)}_{BR} &=\min \{ 1,\frac{\mathcal{N}_{\rm EP}(\beta)}{\mathcal{N}_{\rm BR}(\beta)+1} \cdot \frac{1}{\sqrt{Q}} \} \simeq 1 \;,
\end{align}
where $\mathcal{N}_{\rm IP}(\alpha)+1 = \mathcal{N}_{\rm IP}(\beta) $ and $\mathcal{N}_{\rm BR}(\beta)+1 = \mathcal{N}_{\rm BR}(\alpha)$ are used.
Then, for updates of deleting a bridge bond or adding a backbone bond, more considerations are needed.
For example, after deleting a bridge from configuration $\alpha$, there is no simple relation between $\mathcal{N}_{\rm EP}(\beta)$ and $\mathcal{N}_{\rm EP}(\alpha)$.
Instead, one has $\mathcal{N}_{\rm EP}(\beta)=\mathcal{N}^*_{\rm EP}$, which scales as $\langle \mathcal{N}^*_{\rm EP} \rangle \sim L^{d_{\rm EP} - x_2}$ in the limit $Q \rightarrow 0$ as previously discussed.
For nonzero small $Q$ and large $L$, we expect that Eq.~(\ref{dfbond0}) still holds, thus suppose that $\langle \mathcal{N}^{*}_{\rm EP} \rangle \sim L^2\sqrt{Q}+aL^{d_{\rm EP} - x_2}$.
Substituting this result and Eq.~(\ref{dfbond1}) into Eq.~(\ref{eq:accDeletingBR}),  we get the acceptance probability

\begin{align}
    &P^{(-)}_{BR} = \frac{\mathcal{N}_{\rm BR}}{\mathcal{N}^{*}_{\rm EP}} \cdot \sqrt{Q} \sim \frac{\sqrt{Q}}{\sqrt{Q}+aL^{d_{\rm EP} - x_2 - 2}} \;.
\end{align}

Thus, we expect that the autocorrelation time scales as the inverse of the above acceptance probability 
\begin{align}
    &\tau_{\mathcal{N}_{\rm HF}}/L^2 \sim \frac{\sqrt{Q}+aL^{d_{\rm EP} - x_2 - 2}}{\sqrt{Q}} \;.
\end{align}
From the above equation, for a fixed $Q$, as $L$ increases, since $d_{\rm EP} - x_2 - 2<0$, one has $\tau_{\mathcal{N}_{\rm HF}}/L^2 \to {\cal O}(1)$; 
and for a fixed $L$, as $Q$ decreases, $\tau_{\mathcal{N}_{\rm HF}}/L^2 \to Q^{-1/2}$.
These imply the turning near $Q_c(L)$ and explain the results of the improved Sweeny algorithm, i.e., Fig.~\ref{tauNHF:isweeny}.

\begin{figure}[t] 
\centering
\includegraphics[width=0.5\textwidth]{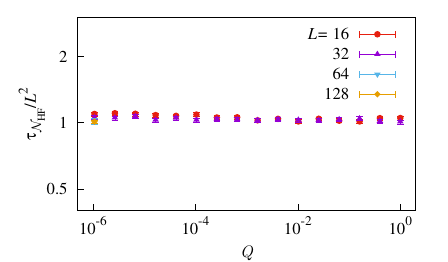}
\caption{
$\tau_{\mathcal{N}_{\rm HF}}$ for the combined method of the improved Sweeny and improved Kawasaki algorithms. 
{The $L$-dependence of $\tau_{\mathcal{N}_{\rm HF}} / L^2$ is very small as demonstrated by data points at $Q=10^{-6}$. Here in our simulation a step consists of a single-bond update in the improved Sweeny algorithm and a two-bond swap in the improved Kawasaki algorithm, which is counted as two ``hits".}
}
\label{isk}
\end{figure}

Finally, since the acceptance probabilities for operations in the improved Kawasaki algorithm are also $1$ for nonzero $Q$, we can combine the improved Sweeny algorithm with the improved Kawasaki algorithm. 
Our simulation results for this combined method are shown in Fig.~\ref{isk},  where $\tau_{\mathcal{N}_{\rm HF}}/L^2$ versus $L$ is almost a straight line. 
Thus the improved Kawasaki method eliminates the $\sim Q^{-1/2}$ form slowing-down of the improved Sweeny algorithm at fixed $L$, and the combined improved Sweeny-Kawasaki method serves as a complete solution for simulating the random-cluster model with small $Q$.

\section{\label{sec:level1}Discussion and conclusion}

To summarize, by introducing the number of half faces, we observe a $Q$-dependent slowing-down in two dimensions, 
which further enriches the dynamics of the Sweeny algorithm for the random-cluster model.
The slowing-down is understood by observing the updating probabilities, which contains the critical bond weight $v_c=\sqrt{Q}$ on the square lattice.
By classifying the bonds into four types and dynamically updating their type information, we propose an improved Sweeny algorithm,
which (at fixed $L$) pushes the slowing-down to much smaller $Q$-region. 
Since the $Q$-dependent slowing-down renders the (improved) Sweeny algorithm being non-ergodic for $Q \rightarrow 0$,
we resort to the Kawasaki algorithm. 
The latter algorithm is found to exhibit an $L$-dependent slowing-down, and this slowing-down can be fully suppressed in an improved version of the algorithm by making use of the bond-type classification.
Finally, we formulate an improved Sweeny-Kawasaki method which combines advantages of the two improved algorithms and {eliminates the above} critical slowing-down.

We note that updates in the improved algorithms proceed with the aid of Baxter-Kelland-Wu loops,
as it is convenient to classify the bonds using the loops, and it is also faster to
perform connectivity checking in the loops than in the clusters.
The computational cost of the algorithms primarily stems from the connectivity checking:
simultaneous breadth-first search~\cite{deng2010some,elcci2013efficient} in the clusters has a runtime scaling of $\sim L^{d_{\rm F}-x_2}$,
while that in the loops has a runtime scaling of $\sim L^{d_{\rm H}-x_2}$ with $d_{\rm H} \leq d_{\rm F}$.
For not too large $L$, the improved Sweeny-Kawasaki method in this work should be adequate for simulating the random-cluster model.
To speedup the simulations for large $L$, it should be worth to study how to incorporate dynamic connectivity algorithms which have runtime scaling at most logarithmic in the system size~\cite{elcci2013efficient,elci2015algorithmic}.

It is also noted that, solely for the uniform spanning trees, there exist algorithms which have excellent performance. 
These include the celebrated Wilson's algorithm~\cite{wilson1996generating}, 
which uniformly constructs a spanning tree from scratch by a loop-erased random walk, 
and {a smart} Kawasaki algorithm {which uniformly generates spanning trees by using a fast link-cut tree data structure~\cite{russo_linking_2018}.
The latter is different from our improved Kawasaki algorithm, which constructs loops to classify the bonds and applies also to models with $Q \ne 0$.}

{Furthermore, though our simulations were performed at $v_c=\sqrt{Q}$ on the square lattice,
for $v=Q^{\sigma}$ with general values of $0 < \sigma <1$, as $Q \rightarrow 0$ there is also 
$Q$-dependent slowing-down for the Sweeny algorithm. This is because in the latter case the updating probabilities tend 
to zero as $Q^{\sigma}$ or $Q^{1-\sigma}$. The improved Sweeny and Kawasaki algorithms can also solve this $Q$-dependent
slowing-down, due to reasons similar to those for $v=\sqrt{Q}$.}
{Also, t}he $Q^{-1/2}$-form critical slowing-down for the random-cluster model is expected to exist on other 2D lattices.
For example, on the triangular ($\mathbb{T}$) and honeycomb ($\mathbb{H}$) lattices, 
when $Q \leq 4$, it is known~\cite{baxter1978triangular,wupotts,grimmett2006random} that 
\begin{align}
    &v^{\mathbb{T}}_c=2\cos[\frac{2}{3}\cos^{-1}(\frac{\sqrt{Q}}{2})]-1 \;,\\
    &v^{\mathbb{H}}_c=2\sqrt{Q}\cos[\frac{1}{3}\cos^{-1}(\frac{\sqrt{Q}}{2})] \;,\\
    &v^{\mathbb{T}}_c \cdot v^{\mathbb{H}}_c=Q \;.
\end{align}
These critical bond weights $v^{\mathbb{T}}_c$ and $v^{\mathbb{H}}_c$ also vanish as $\sim \sqrt{Q}$ when $Q \rightarrow 0$.
However, when the spatial dimension $d \geq 3$, for small $Q$ the critical point behaves as $v_c \sim Q$ (hence the limit $Q \to 0$ corresponds to the spanning-forest model)~\cite{deng2007ferromagnetic}. Thus, the $Q$-dependent critical slowing-down does not occur for $d \geq 3$. Nevertheless, the improved Sweeny and Kawashaki algorithms may still help to improve the simulation efficiency.

{Finally, the number of half faces proves to be very useful in demonstrating the slowing-down of the unimproved algorithms. 
Further work may explore more quantities which have similar behavior. Despite the mechanism discussed in this work, 
we cannot rule out any other property which may cause more severe slowing-down.}

\begin{acknowledgments}
This work has been supported by the National Natural Science Foundation of China (under grant No.~12275263 and No.~12375026), the Innovation Program for Quantum Science and Technology (under grant No.~2021ZD0301900), and the Natural Science Foundation of Fujian Province of China (under grant No.~2023J02032). We thank David B. Wilson for his early contributions on this work.
\end{acknowledgments}

\appendix
\section*{Appendix: Density of bonds in the thermodynamic limit}
In this Appendix, we give an alternative derivation of the results in Eq.~(\ref{rhobonds}).
On the square lattice in the thermodynamic limit $L \rightarrow \infty$, the self-duality condition leads to the following relation between the critical numbers of bonds
\begin{align}
    \langle\mathcal{N}_{\rm BB} + \mathcal{N}_{\rm BR} \rangle/2L^2 = \langle\mathcal{N}_{\rm IP} + \mathcal{N}_{\rm EP} \rangle/2L^2 = 1/2 \;.
\label{eq:half-density}
\end{align}
In the Fortuin-Kasteleyn transformation, the density of occupied bonds can be related to the nearest-neighbor connectivity $g_{\rm n}$ (i.e., the probability that
two nearest-neighbor sites belong to the same cluster) as~\cite{Hu2014}
\begin{align}
    \langle\mathcal{N}_{\rm BB} + \mathcal{N}_{\rm BR} \rangle/2L^2 = p_c [g_{\rm n} + (1-g_{\rm n})/Q] \;,
\label{eq:bonds-gn}
\end{align}
where $p_c = \sqrt{Q}/(1+\sqrt{Q})$.
Substituting Eq.~(\ref{eq:half-density}) into Eq.~(\ref{eq:bonds-gn}), one gets~\cite{Hu2014}
\begin{align}
    g_{\rm n} = \frac{2+\sqrt{Q}}{2(1+\sqrt{Q})} \;.
\label{eq:gn}
\end{align}
Letting $P_{\rm n}$ be the probability that two nearest-neighbor sites are connected by a path not including the bond between the two sites, 
one can express the probability $g_{\rm n}$ as
\begin{align}
    g_{\rm n} = P_{\rm n} + (1-P_{\rm n}) \frac{v/Q}{1+v/Q} \;.
\label{eq:gn-Pn}
\end{align}
Substituting $v_c=\sqrt{Q}$ and Eq.~(\ref{eq:gn}) into Eq.~(\ref{eq:gn-Pn}) leads to $P_{\rm n}=1/2$.
Finally, one can express the critical densities of four kinds of bonds as
\begin{align}
    \langle\mathcal{N}_{\rm BB} \rangle/2L^2 = P_{\rm n} \frac{{v}}{1+{v}} = \frac{\sqrt{Q}}{2(1+\sqrt{Q})} \;,
\end{align}
\begin{align}
    \langle\mathcal{N}_{\rm BR} \rangle/2L^2 = (1-P_{\rm n}) \frac{{v}/Q}{1+{v}/Q} =  \frac{{1}}{2(1+\sqrt{Q})} \;,
\end{align}
\begin{align}
    \langle\mathcal{N}_{\rm IP} \rangle/2L^2 = P_{\rm n} \frac{1}{1+{v}} = \frac{{1}}{2(1+\sqrt{Q})} \;,
\end{align}
\begin{align}
    \langle\mathcal{N}_{\rm EP} \rangle/2L^2 = (1-P_{\rm n}) \frac{1}{1+{v}/Q} = \frac{\sqrt{Q}}{2(1+\sqrt{Q})} \;.
\end{align}

The above densities have been derived in Ref.~\cite{Eren2016}, and they are consistent with $\langle\mathcal{N}_{\rm BB} \rangle/2L^2 = \langle\mathcal{N}_{\rm BR} \rangle/2L^2 = 1/4$ for $Q=1$ in Ref.~\cite{xu2014geometric}.

\bibliographystyle{apsrev4-2}
\bibliography{main}

\end{document}